\documentclass[prb,aps,twocolumn,twocolumngrid,superbib]{revtex4}

\usepackage{graphicx}
\usepackage{amsfonts}
\usepackage{amsmath}
\usepackage{bm}
\usepackage{alltt}
\usepackage{dcolumn} 
\usepackage{graphicx}

\begin{document}

\date{\today}

\title{Linear scaling computation of the Fock matrix VII.  \\
       Periodic Density Functional Theory at the $\Gamma$-point.}

\author{C.~J.~Tymczak}
\affiliation{Theoretical Division, Los Alamos National Laboratory, Los Alamos, New Mexico 87545 }
\author{Matt Challacombe}
\affiliation{Theoretical Division, Los Alamos National Laboratory, Los Alamos, New Mexico 87545 }

\begin{abstract}
Linear scaling quantum chemical methods for Density Functional Theory are extended to the condensed 
phase at the $\Gamma$-point.  For the two-electron Coulomb matrix, this is achieved with a tree-code 
algorithm for fast Coulomb summation [J.~Chem.~Phys.~{\bf 106}, 5526 (1997)], together with  multipole representation 
of the crystal field [J.~Chem.~Phys.~{\bf 107}, 10131 (1997)].   A periodic version of the hierarchical cubature 
algorithm [J.~Chem.~Phys.~{\bf 113}, 10037 (2000)], which builds a telescoping adaptive grid for numerical 
integration of the exchange-correlation matrix, is shown to be efficient when the problem is posed as integration over 
the unit cell.  Commonalities between the Coulomb and exchange-correlation algorithms are discussed, with an emphasis on 
achieving linear scaling through the use of modern data structures.  With these developments, convergence of the 
$\Gamma$-point supercell approximation to the ${\bf k}$-space integration limit is demonstrated for MgO and NaCl. 
Linear scaling construction of the Fockian and control of error is demonstrated for RBLYP/6-21G* diamond up to 512 atoms.\\


\end{abstract}


\maketitle

\footnotetext[1]{\tt tymczak@lanl.gov}
\footnotetext[2]{Preprint LA-UR-03-7723.}

\section{INTRODUCTION}

Quantum chemical methods that  employ Gaussian-Type Atomic Orbitals (GTAOs) 
offer a number of advantages in materials science.  First, because they are local basis functions, it is possible 
to achieve a linear scaling cost with system size for insulating systems.  Secondly,  almost all one- and two-electron 
integrals involving GTAOs are analytic, enabling the rapid evaluation of expectation values involving 
complicated operators that are often involved in the computation of response properties \cite{MHonda91,THelgaker92,PStephens94}.
The {\sc DALTON} quantum chemical program \cite{dalton} is a premier example of this capability, 
offering a wide range of electric and magnetic molecular response properties.   
The ability to treat core-states analytically also opens the ability to go  beyond the pseudopotential 
approximation in computation of relativistic effects with the four-component Dirac-Hartree-Fock \cite{JLaerdahl97,IGrant00} 
and Dirac-Kohn-Sham \cite{TYanai01} theories.  Perhaps most important though, the exact Hartree-Fock (HF) exchange
may be computed efficiently with a GTAO basis set.  In addition to providing a reference for 
correlated wavefunction methods, the exact HF exchange is central to hybrid HF/DFT models \cite{Gill92,Becke93,VBarone96,CAdamo99}.
The use of hybrid methods  in the condensed phase, pioneered by the {\sc CRYSTAL} group \cite{Crystal98},
has proven to be a useful improvement beyond the generalized gradient approximation for a number of properties, including
bulk geometries, electronic properties \cite{TBredow00,JMuscat01} and absorption energies \cite{PBaranek01,AWander01}.

Recently, we have developed linear scaling quantum chemical methods for gas phase Density Functional Theory 
(DFT), including computation of the Coulomb matrix ${\bf J}$ \cite{MChallacombe97} and the exchange-correlation
matrix $\bf K_{\rm xc}$ \cite{MChallacombe00A}.   In this contribution, these linear scaling methods are extended to 
periodic boundary conditions at the $\Gamma$-point.  

With periodic linear scaling quantum chemical algorithms, it is possible to begin bridging the gap between 
methods developed for small molecule chemistry and large scale problems in the solid state.  Together with the 
results presented here, $ {\cal O}(N)$ methods for solving the Self-Consistent-Field equations 
\cite{ANiklasson02A,ANiklasson03} and linear scaling algorithms for computing the periodic HF exchange \cite{CTymczak04b}, 
it is now possible to perform condensed phase  HF/DFT calculations on systems larger than 500 atoms with a single processor.
In addition, with the advent of linear scaling density matrix perturbation theory \cite{ANiklasson04,VWeber04},
well developed quantum chemical methods for the analytic computation of response properties may be brought
to bear on large solid state problems. 

This paper is organized as follows:  In Section~\ref{pbcintro},  periodic boundary conditions and the $\Gamma$-point 
approximation are introduced. Next, in Section \ref{datastruct}, the relationship between the numerical error estimates and data structures that underly the 
fast linear scaling algorithms for computation of the Coulomb and exchange-correlation matrix are outlined.  
In Section~\ref{coulombsums}, we extend previous work on the Niboer and De Wette \cite{BNijboer57,BNijboer58a}  lattice sum 
method to linear scaling computation of quantum Coulomb sums and  tin-foil boundaries.  Then, in Section~\ref{hicu}, 
${\cal O}(N)$ methods for computing the GTAO-based  exchange-correlation matrix are presented. 
In Section~\ref{implement} we discuss the implementation of these developments in the {\sc  MondoSCF} \cite{MondoSCF} suite
of linear scaling quantum chemistry codes.  In Section~\ref{validate}, comparison of the $\Gamma$-point results
is made with those obtained with {\sc CRYSTAL98} using ${\bf k}$-space integration for NaCl and MgO.  Next, 
in Section~\ref{scale}, linear scaling is demonstrated for construction of the diamond Fock matrix at the RBLYP/6-21G*  
level of theory.     Finally, in Section~\ref{conclude}, we present our conclusions. 

\section{periodic boundary conditions, linear scaling and basis sets}\label{pbcintro}

In the conventional implementations of periodic boundary conditions, the 
Bloch functions 
\begin{equation}
\psi^{\bf k}_a({\bf r})  =  \sum_{\bf R} e^{i {\bf k}\cdot {\bf R}} \phi_a ({\bf r}-{\bf R}),
\label{Block}
\end{equation}
are often constructed from non-orthogonal functions local to the unit cell (UC). Here, the 
local function
$\phi_a$ is a Gaussian-Type Atomic Orbital (GTAO) centered on atom {\bf A}, while the 
sum on {\bf R} runs over the Bravais lattice defined by integer translates of the primitive 
lattice vectors {\bf a}, {\bf b} and {\bf  c}.  
These Bloch functions (crystal orbitals) yield all possible translational symmetries through variation of the 
reciprocal lattice vector {\bf k}.   Programs such as {\sc CRYSTAL98} perform a careful sampling of reciprocal space 
to achieve an accurate description of the periodic system.  An alternative approach to including these 
important symmetries is to set ${\bf k}=0$, and then use a larger supercell created through 
replication and translation of the primitive unit cell.  
This is the supercell $\Gamma$-point approximation, used
primarily for the study of defects and vacancies rather than as a replacement for 
${\bf k}$-space integration.

In this contribution, ${\cal O}(N)$ algorithms are developed specifically for the the 
$\Gamma$-point  approximation, allowing the use of large supercells in the case of high symmetry, as well as 
large primary cells in the case of disordered systems.  While ${\bf k}$  dependence is avoided, lattice 
summation and formal integration over the unit cell volume, $V_{\rm UC}$,  are retained.
At first sight this would seem to make matrix construction quite different than in the gas phase,  
where integrals are typically taken over all space, $V_\infty$.  Thus, elements of the gas phase overlap matrix,
\begin{equation}
\label{Sab_norm}
S_{ab}=\int _{V_{\infty }}\, d{\mathbf{r}}\, \phi _{a}({\mathbf{r}})\phi _{b}
({\mathbf{r}}) \, ,
\end{equation}
become 
\begin{equation}
\label{Sab_pbc1}
S_{ab}=\sum _{\mathbf{R},\mathbf{R}'}\int _{V_{\rm UC}}\, d{\mathbf{r}}\, 
\phi _{a}({\mathbf{r}+\mathbf{R}})\phi _{b}({\mathbf{r}+\mathbf{R}'})
\end{equation}
in the periodic $\Gamma$-point regime.  However, this formalism can be  brought into a form 
more closely related to its quantum chemical counterpart via the transformation,
\begin{equation}
\sum_{\bf R} \int_{V_{\rm UC}} d {\bf r} f({\bf r+R}) \rightarrow \int_{V_\infty} d {\bf r} f({\bf r}),
\end{equation}
allowing use of conventional analytic integral technologies.
For example, elements of the periodic overlap matrix become
\begin{equation}
\label{Sab_pbc2}
S_{ab}=\sum _{\mathbf{R}}\int _{V_{\infty }}\, d{\mathbf{r}}\, \phi _{a}
({\mathbf{r}})\phi _{b}({\mathbf{r}+\mathbf{R}}).
\end{equation}

For compactness of notation, let us define the 
intermediate basis function products (distributions)
$\rho _{ab}({\mathbf{r}})=\sum _{\mathbf{R} \mathbf{R}'}\phi _{a}({\mathbf{r}+\mathbf{R}})\phi _{b}({\mathbf{r}+\mathbf{R}'})
$ associated with integration over $V_{\rm UC}$ and the corresponding distributions
$\rho _{ab}^{\scriptscriptstyle \infty}({\mathbf{r}})=\sum _{\mathbf{R}}\phi _{a}({\mathbf{r}})\phi _{b}
({\mathbf{r}+\mathbf{R}})
$ associated with integration over $V_\infty$.
 We likewise define the electron density 
$\rho({\mathbf{r}})=\sum_{ab} P_{ab} \rho _{ab}({\mathbf{r}})
$ associated with integration over $V_{\rm UC}$ and the corresponding 
density $\rho^{\scriptscriptstyle \infty}({\bf{r}})= \sum _{ab} P_{ab} 
\rho _{ab}^{\scriptscriptstyle \infty}({\mathbf{r}})$ associated with integration over $V_\infty$, 
where $P_{ab}$ is the one-electron reduced density matrix.  
In this convention, $V_\infty$ is the default volume of integration, and elements of the periodic overlap 
matrix are expressed simply  as $S_{ab}=\int  d{\mathbf{r}}\, \rho^{\scriptscriptstyle \infty}_{ab}({\mathbf{r}})$,
 while the electron count is
$N_{\rm el} =\int  d{\mathbf{r}}\, \rho^{\scriptscriptstyle \infty}({\mathbf{r}})$.

It is worth noting that the complexity of $\rho^{\scriptscriptstyle \infty}$ is ${\cal O}(N)$, due to the
exponential prefactor $e^{- \chi_{ab} ({\bf A}-{\bf B}-{\bf R})^2}$ that enters each term in the sum 
over {\bf A}, {\bf B} and {\bf R}.  Thus, $N$-scaling may be achieved {\em a priori} with a simple distance test.
However, for small exponents, care must be exercised in  truncation of periodic sums to avoid overlap 
matrices that are not positive definite.  While these situations can often be ameliorated  with a  tighter distance neglect 
criteria, they are typically a symptom of near linear dependence, often  due to the use of basis sets designed for gas phase 
calculations in conjunction with small unit cells.  

These considerations and others are discussed by Towler in an excellent overview of Gaussian basis sets for the condensed 
phase \cite{MTowler00}.  Also, there are at least two (albeit related) libraries \cite{CRYSTALLib,TowlerLib} of Gaussian 
basis sets appropriate for materials at standard densities. For high densities though, these basis sets may 
still encounter problems with linear dependence and sensitivity to truncation.  One solution to this problem, suggested 
by Gr{\"u}neich and Hess \cite{AGruneich98} for even tempered basis sets, is to scale the exponents by the inverse 
square  of the lattice constant.  In many cases though, especially for large systems, standard quantum chemical basis sets 
work well. 


\section{data structures and error estimates}\label{datastruct}

Both the Quantum Chemical Tree-Code (QCTC) \cite{MChallacombe97} 
for computing the Coulomb matrix and Hierarchical Cubature (HiCu) \cite{MChallacombe00A}
for computing the Exchange-Correlation matrix are fast ${\cal O}(N {\rm lg} N)$ algorithms whose performance
is coupled to underlying data structures and error estimates.  It is important to understand some 
of these particulars first, before addressing their extension to periodic boundary conditions.
Also, the current version of QCTC is quite different from previous descriptions, and deserves some introduction.  

Both QCTC and HiCu are homeomorphic, involving $k$-d tree representation of the density.  
In our implementations, $k$-d trees are doubly linked lists with axis aligned bounding boxes (AABBs) 
delimiting the spatial {\em extent} of each node and its children.    This scheme is similar to
well developed technologies for ray tracing and data base searches, allowing fast 
${\cal O}({\rm lg} N)$ {\em range queries} of overlapping components through AABB intersection tests \cite{MGomez99}. 
In the case of QCTC, this fast look up constitutes the penetration acceptability criterion (PAC) 
which identifies spatial clusters or agglomerations, $\rho_Q$, of the density that may be accurately 
represented via a multipole approximation due to the absence of charge-charge penetration effects.  

For accepted clusters a second test, the multipole acceptability criterion (MAC), is performed to 
check translation errors in the multipole  expansion. This second test is critical to the overall accuracy of 
the Coulomb matrix build.  We have recently developed a new MAC in Ref.~[\onlinecite{CTymczak04c}] that  has several 
advantages.  First, it takes into account the magnitude or weight of the distribution within 
the cluster.    Second, it correctly takes into account the angular symmetry  of the primitive Gaussian distributions.   
Third, and most important, it  is always an exact bound to the translation error.   

For each primitive bra distribution $\rho_{ab}$, a fast range query is performed on the $k$-d tree 
representation of the total density, leading to an on the fly partition of near-field (NF) and far-field (FF) 
interactions in construction of the gas phase Coulomb matrix which may be written 
\begin{eqnarray}
J_{ab} &=& \sum_{Q\in \rm FF} \sum_l (-1)^l \sum_m O^l_m[\rho_{ab}] \sum_{l'} \sum_{m'} M^{l+l'}_{m+m'} O^{l'}_{m'}[\rho_Q] 
\nonumber \\
&+& \sum_{q\in \rm NF} \int d {\bf r} \int d {\bf r'} \rho_{ab} ({\bf r}) \left|{\bf r}-{\bf r'} \right|^{-1}
\rho_q({\bf r})  \, ,
\end{eqnarray}
where $M^l_m$ is the irregular solid harmonic interaction tensor, $O^l_m[f]=\int d{\bf r} O^l_m({\bf r}) f({\bf r})$
is a moment of the regular solid harmonic, 
$Q$ runs over the  highest possible nodes in the density-tree consistent with the PAC and MAC, and 
$q$ runs on leftover near-field primitive distributions in the density.
See Refs.~[\onlinecite{MChallacombe97,MChallacombe97D}] for further details on this representation.

A fundamental difference between QCTC and FMM based methods is that QCTC pushes the near/far-field partition to  the limit,  
employing the PAC and MAC best case error estimates to resolve individual primitive distributions.  
On the other hand, FMM based methods employ static, worst case error estimates.  
While recurring down the density tree to the level of individual primitives precludes well developed technologies 
for the integral evaluation of contracted functions, it accelerates the onset of linear scaling through early clustering.

The Quantum Chemical Tree Code generally employs the total density, which simplifies the code, allows 
electrostatic screening in MAC error estimates and provides charge neutrality, an essential
feature for periodic calculations.  Thus, the Coulomb matrix employed here includes the electron-nuclear terms;
${\bf J} \equiv {\bf J}_{\rm ee} + {\bf V}_{\rm en}$.

In the case of HiCu, two separate $k$-d tree structures are used.  The rho-tree holds the electron density,
while the cube-tree contains a hierarchical grid for integration of the exchange-correlation potential.  Each 
node of the 
cube-tree is composed of a Cartesian non-product integration rule with the grid points enclosed by it's AABB.
The cube-tree is constructed iteratively through recursive bisection of the primary volume (the root AABB), using 
exact error bounds to achieve arbitrary precision of the integrated density and its gradients.   As the cube-tree 
is extended, AABB intersection tests are performed while traversing the rho-tree,  avoiding parts of the density 
that do not overlap with that portion of the grid.  Upon construction of the grid,  the reverse procedure is
 carried out;  
for each primitive distribution, the cube-tree is walked selecting only overlapping portions of the grid via the AABB 
intersection test.

For both of these fast algorithms, the trade off between efficiency and accuracy is controlled by the AABB,  
which in turn 
depends on the the extent or range $R_q$ of a primitive Gaussian distribution $\rho_q$, beyond which it is assumed to be
negligible.  Of course, negligible depends on the use to which the distribution is put, as will become 
obvious in the following. 

Both HiCu and QCTC employ the Hermite Gaussian representation of distributions \cite{GAhmadi95}
\begin{equation}
\rho_q ({\bf r}) = \sum_{lmn} d_{lmn} \Lambda^q_{lmn}({\bf r}),
\end{equation}
where 
\begin{equation}
\Lambda^q_{lmn}({\bf r}) = \frac{\partial^{l+m+n} }{\partial Q^l_x \partial Q^m_y \partial Q^n_z} 
e^{-\zeta_q ({\bf r}-{\bf Q})^2}.
\end{equation}
This representation provides an intermediate form into which elements of the density matrix may be folded, and allows
the use of McMurchie-Davidson recurrence relations  \cite{LMcmurchie78} 
in analytic integral evaluation and density evaluation.
For this form,  Cramer's inequality  \cite{EHille26} 
provides a bound for the behavior of a Hermite-Gaussian distribution: 
\begin{equation}
\rho_q ({\bf r})  \leq C_q e^{ -\widetilde \zeta_q ({\bf r -Q})^2 },  
\end{equation}
where
\begin{equation}
C_q=\sum_{lmn} \left|d_{lmn}\right| K^3 \left[ 2^{l+m+n} ~ l! m! n! ~ \zeta^{l+m+n}_q \right]^{1/2} ,
\end{equation}
the constant $K=1.09$, and  
\begin{equation}
\widetilde \zeta_q = 
\begin{cases}
\zeta_q   & l+m+n= 0 \\
\frac{1}{2} \zeta_q   & {\rm otherwise.}
\end{cases}
\end{equation}

The overlap extent $R^{\rm o}_q$ is the value beyond which numerical evaluation of the 
distribution $\rho_q$ yields a value less than $\tau$;
\begin{equation}
C_q e^{ -\widetilde \zeta_q (R^{\rm o} _q)^2 }  = \tau . \label{OverlapExtent}
\end{equation}
For QCTC,  errors in the electrostatic potential due to penetration errors must be considered.
For this purpose, the penetration extent $R^{\rm p}_q$ is introduced, satisfying the equation 
\begin{equation}
\label{PotentialExtent}
C_q \int \left[ \left(\pi/{\widetilde \zeta_q} \right)^{3/2}  
\delta({ r})  - e^{ -\widetilde \zeta_q {r}^2 } \right] \left|{ r}-{ R^{\rm p}_q} \right|^{-1} {\rm d} r = \tau .
\end{equation}

In both HiCu and QCTC, the density-tree is constructed by recursively splitting the largest
box dimension, until each primitive has been resolved.  Then the primitive AABBs are computed from their
extents and merged recursively back up the tree.  For HiCu, this is all there is to it, but for QCTC multipole
moments are also translated to a common center and recursively merged up the tree.  Also, when computing 
matrix elements of $\bf J$, the primitive bra AABB is computed with $R^{\rm o}_q$, while the  $R^{\rm p}_q$ are 
used to construct AABBs of the density-tree.

In Fig.~1, differences between the penetration and overlap extent are shown for a diffuse $s$-type
Gaussian.  For large extents, such as those encountered in a static FMM-type error bound, the two extents behave
in a similar way.  However, with aggressive use of the multipole approximation as in QCTC, the distinction becomes
critical. 

%
%
\begin{figure}
\includegraphics{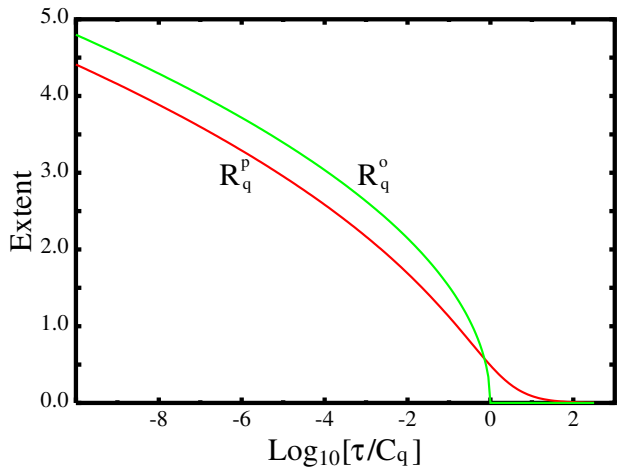}
\caption{Behavior of the overlap extent $R^o$ and the penetration extent $R^p$ 
	 as a function of $\tau/C_q$ for an $s$-type Gaussian with exponent $\zeta=1$. 
         For small $C_q$ (occurring for example due to a large atom-atom separation and/or small density 
         matrix prefactor), $R^o$ goes to zero at the origin and its distribution is eliminated, 
         while $R^p$ goes slowly to zero due to the Coulomb singularity.}
\label{extent}
\end{figure}

\section{Periodic Quantum Coulomb Sums}\label{coulombsums}

In the $\Gamma$-point approximation, elements of the periodic Coulomb matrix are 
\begin{eqnarray}
J_{ab}&=&\int _{V_{\rm UC}} d{\mathbf{r}}  \int 
d{\mathbf{r}'}{\rho _{ab}\left( {\mathbf{r}}\right) {\left| \mathbf{r}-\mathbf{r}'\right| }^{-1}
\rho_{\rm tot}\left( \mathbf{r}'\right) } \\
&=&\sum _{\mathbf{R}}\int\int
d{\mathbf{r}}\, d{\mathbf{r}'}{\rho _{ab}^{\scriptscriptstyle \infty}\left( {\mathbf{r}}\right)  
{\left| \mathbf{r}-\mathbf{r}'\right| }^{-1} \rho^{\scriptscriptstyle \infty}_{\rm tot}
\left( {\mathbf{r}'}+{\mathbf{R}}\right) }
\nonumber 
\end{eqnarray}
where $\rho_{\rm tot}$ is the total, periodic density including both electronic and nuclear terms.  These integrals
involve infinite summation over the lattice vectors ${\bf R}$, and must be handled with care.  There are at least two
main approaches to handling this summation:  Multipole expansion of the Ewald potential 
or Ewald-like summation of the multipole expansion. Expansion of the Ewald potential yields  
tin foil (TF) boundary conditions, requires reciprocal and real space summation with every ${\bf J}$ build, 
and scales as $O(N^{3/2})$.  An alternative is the Ewald-like summation of the multipole interaction tensor,
which was first described by Nijboer and De Wette (NDW) \cite{BNijboer57,BNijboer58a} and later reviewed and extended 
by Challacombe, White and Head-Gordon \cite{MChallacombe97D} to lattice summation of the irregular solid harmonic 
multipole interaction tensor. This Ewald-like summation is taken over the periodic far field, $V_{\rm PFF}$, and is
equivalent to a  direct lattice summation (not a true Ewald sum) excluding an inner region, $V_{\rm In}$, surrounding 
the unit cell.  This inner region has been subtracted to avoid penetration errors and to guarantee convergence of the 
multipole expansion. With the summed interaction tensors cheaply precomputed and reused, the cost of Coulomb 
summation over the PFF scales as as ${\cal O}(N p^2)$, where $p$ is the order of the multipole expansion.  With this 
partition,  the  $N$-scaling periodic quantum Coulomb sums 
involve the contributions
\begin{equation}
{\bf J} = {\bf J}^{\rm In} + {\bf J}^{\rm PFF} + {\bf J}^{\rm TF} \, ,
\end{equation}
corresponding to the three separate regions shown in Fig.~2.   Here, ${\bf J}^{\rm In}$ is computed using 
the fast ${\cal O}(N {\rm lg} N)$ QCTC algorithm outlined previously in Section~\ref{datastruct}. Construction of 
${\bf J}^{\rm PFF}$ will be developed in the following section, while in Section~\ref{tinfoil} the term 
${\bf J}^{\rm TF}$, necessary to 
introduce tin-foil boundary conditions, is detailed.
%
%
\begin{figure}
\resizebox*{3.5in}{!}{\includegraphics{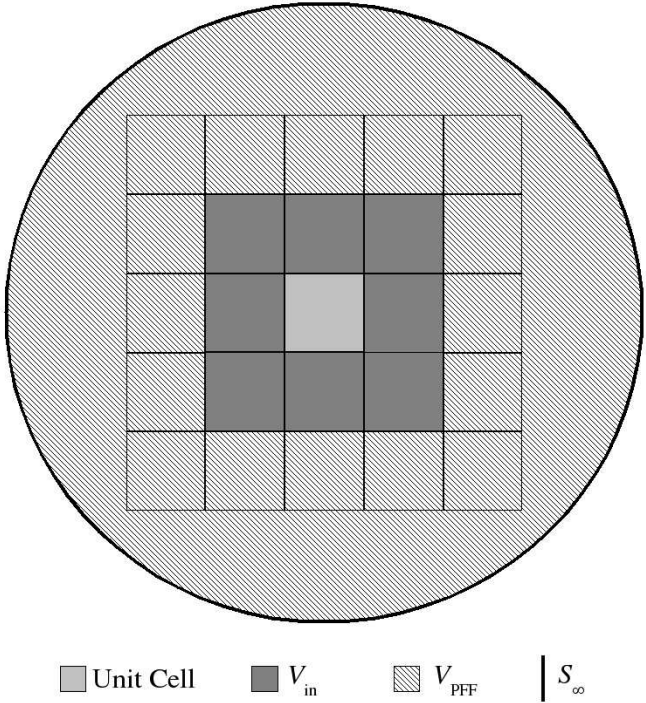}}
\caption{Schematic of the regions contributing to $N$-scaling summation of the Coulomb matrix.
The inner cells that make up $V_{\rm In}$ provide a buffer region that guarantees convergence of the multipole 
expansion of Coulomb interactions between the unit cell and all cells in $V_{\rm PFF}$.  The
periodic far-field region, $V_{\rm PFF}$, is the spherically ordered lattice extending
to infinity but excluding $V_{\rm In}$.  For large cells and/or high order multipole 
expansions, $V_{\rm In}$ includes just the unit cell's 27 nearest neighbors.  In FMM  notation this corresponds to ws=1.  
However, for smaller cells and/or lower order multipole expansions, $V_{\rm In}$ tends to a spherical distribution of 
cells surrounding the unit cell.   Direct summation over $V_{\rm PFF}$ leads to charges at the infinite surface  
$S_{\infty}$, which must be be canceled by tin-foil (conducting) boundary conditions to achieve equivalence with 
Ewald summation.}
\label{regions}
\end{figure}

\subsection{The Periodic Far Field}\label{pff}

By construction, the periodic far field (PFF) term in the Coulomb matrix,
\begin{equation}
J_{ab}^{\rm PFF} = \sum _{{\bf R}\in {\rm PFF}} \int \int d {\bf r} d {\bf r'}  \rho^\infty_{ab}
({\bf r}) {\left| {\bf r}-{\bf r'}+{\bf R} \right|}^{-1}
\rho^\infty_{\rm tot} ({\bf r'}),
\label{PFFOne}
\end{equation}
involves charge distributions that are well separated with respect to both penetration and the
convergence of multipole expansion errors, as outlined in Fig.~2 and discussed in the following. 

With these conditions, and assuming the unit cell is centered at the origin, the bipolar multipole expansion 
employing the regular and irregular solid harmonics, $O^l_m$ and $M^l_m$ respectively is 
\begin{eqnarray}
\label{bipolar}
{| {\bf r}-{\bf r}'+{\bf R}|}^{-1} &\approx&
\sum^p_{l=0} (-1)^{l}  \,
\sum^p_{l'=0} \Bigl[
\\
&&\sum^{l}_{m=-l} \,
\sum^{l'}_{m=-l'} 
O^{m}_{l}({\bf r})
            \,M^{m+m'}_{l+l'}({\bf R})
            \,O^{m'}_{l'}({\bf r}')  \Bigr] \, . \nonumber
\end{eqnarray}
Inserting Eq.~(\ref{bipolar}) into into Eq.~(\ref{PFFOne}) yields
\begin{eqnarray}
J_{ab}^{\rm PFF}& =&\sum _{{\bf R}\in {\rm PFF}} \sum^p_{l=0} (-1)^{l}  \, \sum^p_{l'=0} \sum^{l}_{m=-l} 
\Bigl(  \\
&&
\sum^{l'}_{m=-l'} 
O_{l}^{m}\left[ \rho^\infty_{ab}\right]\,M^{m+m'}_{l+l'}({\bf R}) \, O_{l'}^{m'}\left[ \rho^\infty_{\rm tot}\right] 
\Bigr) \, .  \nonumber
\end{eqnarray}
This expression decouples the complexity of $\rho^\infty_{ab}$ from $\rho^\infty_{\rm tot}$
through the precomputed multipole moments
$O_{l}^{m}\left[ \rho^\infty_{ab}\right] 
=\int  d\mathbf{r}\, {O}_{l}^{m} \left( \mathbf{r}\right) \, \rho^\infty_{ab} \left( \mathbf{r}\right) $
and 
$O_{l}^{m}\left[ \rho^\infty_{\rm tot}\right] =\int  d\mathbf{r}\, {O}_{l}^{m} \left( \mathbf{r}\right) \, 
\rho^\infty_{\rm tot} \left( \mathbf{r}\right)$.
Following Nijboer and De Wette \cite{BNijboer57,BNijboer58a,MChallacombe97D}, we introduce the effective multipole 
interaction tensor 
\begin{equation}
{\cal M}^{m}_l=\sum_{{\bf R}\in V_{\rm PFF} } M^m_l({\bf R}) \;,
\end{equation}
which can be efficiently computed on the fly for each new lattice, both to high accuracy and to high order (large $p$) 
using the new methods detailed in Appendix~\ref{calMTen}.  Note that this is a direct sum of the interaction tensor, 
and is {\em not} equivalent to Ewald summation.  Nevertheless, with this simplification,  the ${\cal O}(p^2 N)$  
working equation 
\begin{equation}
\label{WorkingPFF}
J_{ab}^{\rm PFF}= \sum^p_{l=0} \sum^{l}_{m=-l} O_{l}^{m}\left[ \rho^\infty_{ab}\right] {\cal J}^m_l \;,
\end{equation}
is obtained, where the intermediate tensor 
\begin{equation}
{\cal J}^l_m = (-1)^{l}\sum^p_{l'=0} \sum^{l'}_{m=-l'}  {\cal M}^{m+m'}_{l+l'} O_{l'}^{m'}
\left[ \rho^\infty_{\rm tot}\right] 
\end{equation}
is cheaply precomputed at the start of each Coulomb build.

Because Eq.~(\ref{WorkingPFF}) is inexpensive, our strategy is to define a minimal buffer region, $V_{\rm In}$, sufficient 
to control penetration errors, subtracting effort from the computation of $J^{\rm In}$ via QCTC and replacing
it with cheaper, multipole work in the computation of $J^{\rm PFF}$.  To this end, a fixed inner region $V_{\rm In}$ 
is constructed 
from neighboring cells that have simple Gaussian overlap with the unit cell, defined by the radius $R^{o}$.  
As explained in Section~\ref{datastruct}, for the relatively large distances considered at this level the differences 
between the
penetration and overlap extent are negligible. 
With $V_{\rm In}$ fixed, the precision of ${\bf J}^{\rm PFF}$ is controlled entirely by the expansion order $p$.  
In general $p$ will be much higher than the expansion order ($\sim 5$) employed by QCTC in computation of 
${\bf J}^{\rm In}$.  With QCTC accuracy is controlled on the fly by the MAC and PAC, establishing a dynamic 
near/far-field partition, while computation of ${\bf J}^{\rm PFF}$ involves a static, worst-case error dominated by 
the multipole expansion.  
This static error is controlled by using the FMM-like error bound, 
\begin{equation}
\label{FMMError}
\frac{ 2^{p} {\cal C}^2 \,{d}_{\rm max}^{p+1}}
{({R}^o)^{p+1}\left| {R}^o -2 \, {d}_{\rm max} \right| } 
\leq \tau_{\rm \scriptscriptstyle MAC} \;,
\end{equation}
to set the appropriate expansion order $p$.  In Eq.~(\ref{FMMError}),  $d_{\rm max}$ is the maximum translational distance, 
${\cal C}$ is the asymptotic Uns{\"o}ld weight of the total density and $\tau_{\rm \scriptscriptstyle MAC}$ is the threshold controlling the 
translation errors. See Ref.~\onlinecite{CTymczak04c} for development of this expression  and further explanation of these parameters.

\subsection{Tin-Foil Boundary Conditions}\label{tinfoil}

The surface charges created by direct summation over $V_{\rm PFF}$ must be canceled
to achieve equivalence with Ewald summation.   Achieving this equivalence is more than
semantic, since without tin-foil boundary conditions matrix elements lack translational
invariance and often incur dramatic charge sloshing instabilities.   The correction is 
strongly dependent on ordering of the direct sum; as the Nijboer and De Wette 
method corresponds to spherical summation due to symmetry of the real/reciprocal space
partition, the appropriate correction is \cite{Redlack72} 
\begin{equation}
\Phi _{\rm Ew}\left( \mathbf{r}\right) =\Phi _{\rm SS}\left( \mathbf{r}\right) + 
\frac{2\pi }
{3V_{\rm UC}} \left(  Q - 2 \, {\bf r } \cdot \mathbf{D} \right),
\label{EW_pot}
\end{equation}
where ${\bf D}$ is the system dipole moment, $Q$ is trace of the system quadrupole and 
we have assumed origin centering, 
The tin-foil correction to the Coulomb matrix is then 
\begin{equation}
J_{ab}^{TF}  = \frac{2\pi }{3V_{\rm UC}} \left(  Q\, S_{ab}-2 \, \mathbf{d}_{ab}\cdot 
\mathbf{D} \right) \label{JTC}
\end{equation}
with  $S_{ab}$ an element of the overlap matrix, and ${\bf d}_{ab}$ the
dipole moment of the distribution $\rho^\infty_{ab}$.  

%
%
\begin{figure} 
\resizebox*{3.5in}{!}{\includegraphics{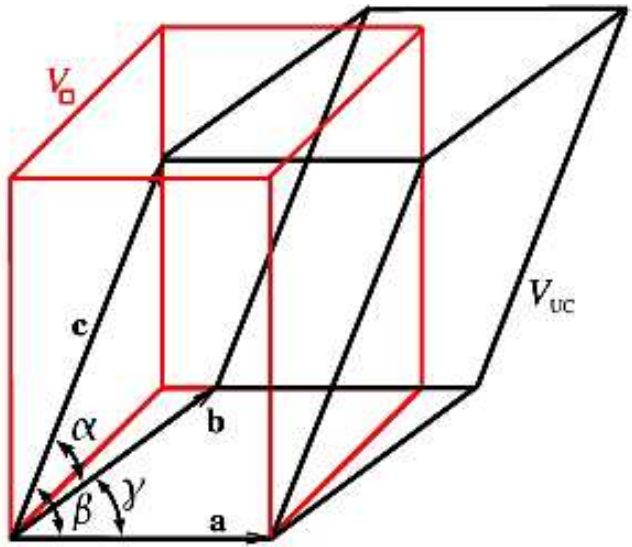}}
\caption{Transformation between  the unit cell with volume  $V_{\rm UC}$ (described by the primitive lattice vectors 
{\bf a}, {\bf b} and {\bf c}) 
and the rectangular integration volume $V_\Box$ employed by HiCu.}
\label{boxuc}
\end{figure}

\section{Periodic exchange-correlation} \label{hicu}

The HiCu algorithm is ideally suited for periodic boundary conditions, 
as the unit-cell $V_{\rm UC}$ can be simply transformed into an equivalent rectangular 
integration domain $V_\Box$ that is the cube-tree's AABB.  These volumes,
shown in Fig.~3,  are equivalent due to full periodicity of both  distributions 
and density. The integration is then simply
\begin{equation}
K^{\rm xc}_{ab} = \int _{V_{\Box}}\, d\mathbf{r}\, \rho_{ab}({\bf r}) v_{\rm \scriptscriptstyle xc}\left[\rho;{\bf r} \right]
\label{Kxc_pbc_0}
\end{equation}
This approach should be contrasted with more conventional quantum chemical methods for computing the 
exchange-correlation matrix, involving the ``Becke weights'' \cite{ABecke88}, which demand numerical integration 
over $V_\infty$.  

While we have written Eq.~\ref{Kxc_pbc_0} in terms of the exchange-correlation potential for simplicity, in practice 
HiCu employs the Pople, Gill and Johnson formulation \cite{JPople92,MChallacombe00A}. 

Because the distributions and density both involve a double sum over lattice vectors, 
there will be a large number of atom-atom pairs that do not overlap with $V_\Box$.  A similar situation is 
encountered in the gas phase for parallel versions of HiCu \cite{CGan03}, where each processor has a small, 
local cube-tree that may overlap only a few of the many possible atom-atom pairs.  
The solution to this problem  again comes from the ray tracing literature, in the form of a modified ray-AABB \cite{MGomez99} and sphere-AABB test \cite{GGems}.  
The ray-AABB test has been modified into a cylinder-AABB test, where the radius of the cylinder
is a maximal overlap extent of the atom-atom pair.  In the case of a same center atom-atom pair,
it is of course more appropriate to employ a sphere-AABB test.  In both cases, overlap
between the HiCu integration volume and atom-atom pairs is established with a negligible prefactor when using these tests.

\begin{table}
\caption{Comparison of {\sc CRYSTAL98} and {\sc MondoSCF} $\Gamma$-point calculations on  NaCl at the 
RBLYP/8-511/8-631G level of theory.  }
\label{NaCl8511}
\center{
\begin{tabular}{lcll}
\toprule
Program         & $N_{\rm at}$              & Energy (au)  & Energy/$N_{\rm at}$\\ 
\colrule
{\sc MondoSCF}      & $2^f$  &  -622.39101   & -311.19551  \\
{\sc CRYSTAL98}     & $2^f$  &  -622.39114    & -311.19557    \\
{\sc MondoSCF}      & $8^g$  &  -2490.0016   & -311.25020  \\
{\sc CRYSTAL98}     & $8^g$  &  -2490.0013   & -311.25016   \\
\botrule
$ ^f$Triclinic \\
$ ^g$Cubic
\end{tabular}}
\end{table}

\begin{table}
\caption{Convergence of the $\Gamma$-point super-cell approximation for NaCl, computed with {\sc MondoSCF} 
         at the RLDA/STO-3G level of theory.}
\label{NaClGammaPt}
\center{
\begin{tabular}{lcll}
\toprule
Program         & $N_{\rm at}$              & Energy (au) & Energy/$N_{\rm at}$\\ 
\colrule
{\sc MondoSCF}  & 2$^f$    &   -610.97536  & -305.48768   \\
                & 8$^g$    &  -2444.3584   & -305.54480   \\
                & 16$^f$   &  -4888.7002   & -305.54377   \\
                & 54$^f$   & -16499.490    & -305.54611   \\
                & 64$^g$   & -19554.956    & -305.54618   \\
                & 128$^f$  & -39109.912    & -305.54618   \\
                & 216$^g$  & -65997.977    & -305.54619   \\ 
\hline
{\sc CRYSTAL98}$^h$ & 2$^f$    &  -611.09228    & -305.54614    \\ 
\botrule
$ ^f$Triclinic \\
$ ^g$Cubic  \\
\multicolumn{4}{l}{$ ^h6\times6\times6$ $\bf k$-space grid.}
\end{tabular}}
\end{table}

\newpage
\newpage

\begin{table}
\caption{Convergence of the $\Gamma$-point super-cell approximation for MgO, computed with {\sc MondoSCF} 
          at the RBLYP/8-61G/8-51G level of theory.}
\label{MgOGammaPt}
\center{
\begin{tabular}{lrll}
\toprule
Program         & $N_{\rm at}$              & Energy (au) & Energy/$N_{\rm at}$\\ 
\colrule
{\sc MondoSCF}  & 2$^f$    &  -275.09097  & -137.54548  \\
                & 8$^g$    &  -1101.7295  & -137.71618  \\
                & 16$^f$   &  -2203.6904  & -137.73065  \\
                & 54$^f$   &  -7437.7989  & -137.73702  \\
                & 64$^g$   &  -8815.2131  & -137.73771  \\
                & 128$^f$  &  -17630.430  & -137.73774  \\
                & 216$^g$  &  -29751.352  & -137.73774  \\ 
\hline
{\sc CRYSTAL98}$^h$ & 2$^f$    &   -275.47547  & -137.73774  \\ 
\botrule
$ ^f$Triclinic \\
$ ^g$Cubic  \\
\multicolumn{4}{l}{$ ^h6\times6\times6$ $\bf k$-space grid.}
\end{tabular}}
\end{table}

\begin{table}
\caption{Convergence of the $\Gamma$-point super-cell approximation for diamond, 
         computed with {\sc MondoSCF} at the RBLYP/6-21G* level of theory.}
\label{DiamondGammaPt}
\center{
\begin{tabular}{rll}
\toprule
 $N_{\rm at}$  & Energy (au) & Energy/$N_{\rm at}$\\ 
\colrule
  8    &   -303.989 &  -37.9986 \\
  16   &   -608.667 &  -38.0417 \\
  32   &   -1218.02 &  -38.0632 \\
  64   &   -2436.28 &  -38.0669 \\
  96   &   -3654.59 &  -38.0687 \\
 144   &   -5482.04 &  -38.0697 \\
 216   &   -8223.10 &  -38.0699 \\
 288   &   -10964.1 &  -38.0700 \\
 384   &   -14618.9 &  -38.0700 \\ 
\botrule
\end{tabular}}
\end{table}

\section{Results}

\subsection{Implementation}\label{implement}

All developments were implemented in a serial version of {\sc MondoSCF} v1.0$\alpha$9 \cite{MondoSCF}, a suite of 
linear scaling Quantum Chemistry code.  The code was compiled using the Portland 
Group F90 compiler {\sc pgf90} v4.2 \cite{pgf90} with the {\tt -O1 -tp athlon} options  and with the 
Gnu C compiler {\sc gcc} v3.2.2 using the {\tt -O1} flag.  All calculations were carried out on a 
1.6GHz AMD Athlon running RedHat  {\sc Linux}v9.0 \cite{RedHat90}.   

Thresholds controlling the cost to accuracy ratio of HiCu and QCTC
are set by the accuracy levels {\tt LOOSE}, {\tt GOOD} and {\tt TIGHT},
which have been empirically chosen to deliver 4-5, 6-7 and 8-9 digits,  respectively,
of relative accuracy in the energy.  Values of these thresholds are listed in Appendix B of Ref.~[\onlinecite{CTymczak04b}].
The unmodified two-electron threshold $\tau_{\rm \scriptscriptstyle 2E}$ sets the 
overlap extent $R^o_p$ in Eq.~(\ref{OverlapExtent}) and the penetration extent $R^p_q$ in Eq.~(\ref{PotentialExtent}),
both of which control the PAC.  As explained in Ref.~[\onlinecite{CTymczak04c}], the threshold $\tau_{\rm \scriptscriptstyle MAC}$
controlling the MAC is set as $\tau_{\rm \scriptscriptstyle MAC}=10^2 \tau_{\rm \scriptscriptstyle 2E}$.  
The HiCu threshold $\tau_{\rm \scriptscriptstyle HICU}$ likewise sets two sub-thresholds.  The
overlap extent $R^o_p$ in Eq.~(\ref{OverlapExtent}), defining accuracy of the density on the grid, is set using 
$10^{-1} \tau_{\rm \scriptscriptstyle HICU}$ ($\tau_\rho$ in Ref.~[\onlinecite{MChallacombe00A}]). The
target relative error defining accuracy of the HiCu grid is just $\tau_{\rm \scriptscriptstyle HICU}$ 
($\tau_r$ in Ref.~[\onlinecite{MChallacombe00A}]).  
It should be pointed out that of all the thresholding schemes, those governing HiCu are the least conservative;
it is a simple (and not too expensive) matter to simply tighten the HiCu threshold if intermediate accuracies
are required. 

The multipole interaction and contraction code used by QCTC in the near/far-field partition has been 
highly optimized by symbolic manipulation and factorization, using real arithmetic and expansions through
7'th order in the calculation of ${\bf J}^{\rm In}$. The computation of ${\bf J}^{\rm PFF}$ 
employs a general code for multipole contraction, allowing expansion through $p=64$.
Eigensolution of the self-consistent-field equations has been used throughout, with the corresponding
matrix and distribution thresholds given Appendix B of Ref.~[\onlinecite{CTymczak04b}].
All calculations were performed with $C_1$ (no) symmetry, and all results are reported in atomic units.
%
%
\begin{figure}
\includegraphics{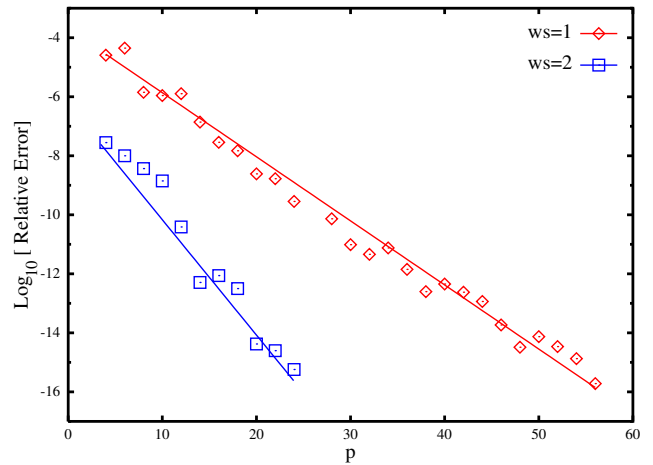}
\caption{Error in the Coulomb energy computed with the 
Nijboer and De Wette scheme relative to true Ewald summation.  Shown
is the error in the Coulomb energy versus $p$ with one (ws=1) and two (ws=2) layers of
cells in $V_{\rm PFF}$ for a periodic system of 64 classical water molecules.}
\label{ErrrorPFF}
\end{figure}

\subsection{Validation}\label{validate}

The ability of our implementation to reproduce true Ewald summation is shown in 
Fig.~4 for a periodic system of 64 classical water molecules.
Note that with both the Ewald sum and the Nijboer and De Wette approach,  ordering the
real and reciprocal space sums is critical; high order agreement is achieved only when
the summation proceeds from the smallest to the largest terms.

The use of Cartesian Gaussian basis sets in many cases allows direct numerical 
comparison of different programs, at least to within the approximations, grids, etc 
peculiar to a code. Here, we make connection with the preeminent Gaussian orbital 
program for periodic calculations, {\sc CRYSTAL98} \cite{Crystal98}.  Calculations have been carried 
out largely with basis sets optimized for the condensed phase\cite{CRYSTALLib}, which tend to 
have less diffuse valence functions.  Tables~\ref{NaCl8511}-\ref{MgOGammaPt} make a direct comparison 
with {\sc CRYSTAL98} for NaCl and MgO obtained with the {\sc MondoSCF} {\tt TIGHT} precision level. 
For the {\sc CRYSTAL98} calculations, we used the following threshold parameters: 
{\tt TOLDENS=10}, {\tt TOLPOT=10}, {\tt TOLGRID=15} and {\tt BASIS=4}.   The {\tt BASIS} parameter determines 
the auxiliary functions used to fit the exchange-correlation potential 

In Table~\ref{NaCl8511}, comparison is made for $\Gamma$-point NaCl with the 8-511G\cite{CBS:8511G:Na} basis set
for sodium, the 8-631G\cite{CBS:8631G:Cl} basis set for chlorine and using the restricted BLYP functional 
\cite{DFT_LYP:88,BLYP:88}.
Next, in Table~\ref{NaClGammaPt}, convergence of the supercell $\Gamma$-point approximation 
is demonstrated for NaCl with the STO-3G basis set and the restricted local density approximation.
Then, in Table~\ref{MgOGammaPt}, convergence of the supercell $\Gamma$-point approximation
to the  ${\bf k}$-space integration result is demonstrated for MgO, using the 8-61G\cite{CBS:861G:MgO} basis set for 
magnesium, the 8-51G\cite{CBS:861G:MgO} basis set for the oxygen, and the restricted BLYP functional.
The primitive lattice coordinates for these systems are given in Ref.~[\onlinecite{PBCCoordinates}].

Finally, in Table~\ref{DiamondGammaPt}, convergence of the supercell $\Gamma$-point approximation is shown for
diamond at the {\tt GOOD} accuracy level,  using the restricted BLYP functional and the
6-21G*\cite{CBS:621G:C} basis set.  Since {\sc MondoSCF} employs 6-$d$ and 10-$f$ functions, while {\sc CRYSTAL98} 
employs 5-$d$ and 7-$f$, we were not able to make a direct comparison for this basis set.

\subsection{Scaling and accuracy}\label{scale}
%
%
\begin{figure}
\includegraphics{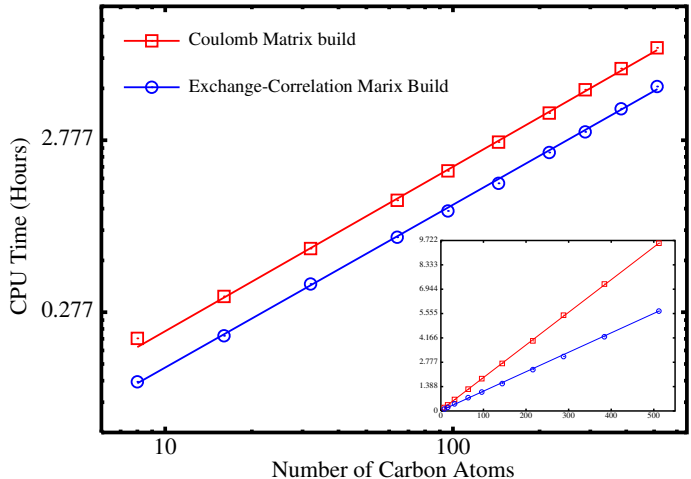} 
\caption{Computational complexity of the ${\bf J}$ and ${\bf K}_{\rm xc}$ matrix builds for 
cubic diamond at the RBLYP/6-21G* level of theory at the {\tt GOOD} accuracy level.}
\label{Scaling_Matrix_Build}
\end{figure}
%
%
\begin{figure}
\includegraphics{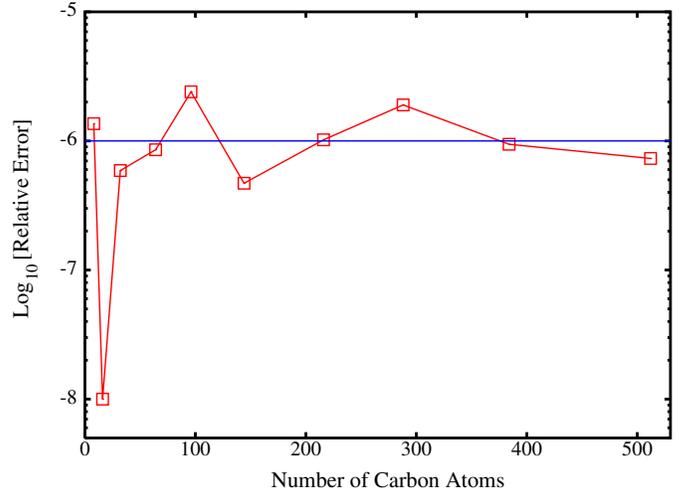}
\caption{Relative error with system size for RBLYP/6-21G* diamond at the {\tt GOOD} accuracy level.}
\label{ErrorPerN} 
\end{figure}

Demonstrating linear scaling at the outset,  Fig.~5 shows the CPU time for ${\bf J}$ and ${\bf K}_{\rm xc}$
builds with  RBLYP/6-21G* diamond at its standard density.  These timings correspond to a {\tt GOOD} level of accuracy, 
targeting 6 digits in the total energy and corresponding to the values listed in Table\ref{DiamondGammaPt}.
Shown in Fig.~6 is the precision of the computed energies, obtained by performing a second set 
of calculations with all thresholds reduced by one order of magnitude.  For these calculations, the largest 
source of error is the numerical integration performed by HiCu, as the QCTC thresholds are significantly more conservative. 

\section{Conclusions}\label{conclude}

We have extended linear scaling quantum chemical methods for computation of exchange-correlation and Coulomb matrices to
periodic boundary conditions at the $\Gamma$-point.  These methods have demonstrated an early onset of linear scaling and
error control for diamond, allowing calculations up to 512 atoms at the RBLYP/6-21G* level of theory.  In both cases, this 
early onset of linear scaling has been enabled by the use of modern data structures such as the $k$-d tree, together with 
reliable error estimates for the Gaussian extent.  These algorithms have been parallelized \cite{CGan03,CGan04B}, demonstrating
high efficiencies up to 128 processors, and have been used recently to determine the $T=0$K equation of state for 
pentaerythritol tetranitrate \cite{CGan04A} at the RPBE/6-31G** level of theory and  a {\tt GOOD} accuracy level.

While this contribution has focused on demonstrating linear scaling for diamond, 
the methods presented here work for slabs and wires as well, using methods for computation 
of the two- and one-dimensional the $\cal M$ tensor outlined in Appendix \ref{calMTen}.
Our experience has shown that, for the same number of atoms, these lower dimensional systems 
run much faster.  

\section*{ACKNOWLEDGMENTS}

We would like to acknowledge Tommy Sewell and Ed Kober for their advice
and support. We would also like to thank Chee Kwan Gan for a careful reading of this manuscript. 

\bibliographystyle{apsrmp} 
\bibliography{pbcdft}

\newpage

\appendix

\section{Computation of the \protect\( {\cal M}\protect \) Tensor}\label{calMTen}

Following Nijboer and De Wette \cite{BNijboer57,BNijboer58a}, and later 
Challacombe, White and Head-Gordon \cite{MChallacombe97D} (CWHG), we begin with 
the partition
\begin{equation}
\label{C1}
\frac{1}{r^{l+1}}={\cal G}_{l}\left( \beta ,r\right) +{\cal F}_{l}\left( \beta ,r\right)
\end{equation}
involving the functions 
\begin{equation}
\label{C2}
{\cal G}_{l}\left( \beta ,r\right) =\frac{\Gamma \left( l+\frac{1}{2},\beta ^{2}r^{2}\right) }
{\Gamma \left( l+\frac{1}{2}\right) \, r^{l+1}}
\end{equation}
and 
\begin{equation}
\label{C3}
{\cal F}_{l}\left( \beta ,r\right) =\frac{\gamma \left( l+\frac{1}{2},\beta ^{2}r^{2}\right) }
{\Gamma \left( l+\frac{1}{2}\right) \, r^{l+1}} \, ,
\end{equation}
where $\Gamma$ is the gamma function,  $\gamma$ is the incomplete gamma function \cite{MAbramowitz87} and \( \beta  \) is a parameter controlling the 
partition.  With this separation of length scales, the lattice sum defining the multipole interaction tensor, ${\cal M}^{m}_{l}$, 
may be expressed as 
\begin{eqnarray}
{\cal M}^{m}_{l} & = & \sum _{\mathbf{R}\in V_{\rm PFF}}\, M_{l}^{m}[\mathbf{R}]
\nonumber\\
 & = & \sum _{\mathbf{R}\in V_{\rm PFF}}\widetilde{P}_{l}^{m}\left( \cos \theta _{\mathbf{R}}
\right) e^{im\phi _{\mathbf{R}}}\,  {\cal G}_{l}\left( \beta ,\left| \mathbf{R}\right| \right)
\nonumber\\
 & + & \sum _{\mathbf{R}\in V_{\rm PFF}}\widetilde{P}_{l}^{m}\left( \cos \theta _{\mathbf{R}}
\right) e^{im\phi _{\mathbf{R}}}\, 
{\cal F}_{l}\left( \beta ,\left| \mathbf{R}\right| \right) \;.
\nonumber\\
\label{C4a0}
\end{eqnarray}
Following  CWHG, this expression can be further developed into real and reciprocal space terms:
\begin{eqnarray}
{\cal M}^{m}_{l} &=&\sum _{\mathbf{R}\in V_{\rm PFF}}\, \widetilde{P}_{l}^{m}
\left( \cos \theta _{\mathbf{R}}\right) e^{im\phi _{\mathbf{R}}}\, {\cal G}_{l}\left( \beta ,\left|
 \mathbf{R}\right| \right) 
\nonumber\\
&-&\sum _{\mathbf{R}\in V_{\rm In}}\widetilde{P}_{l}^{m}\left( \cos 
\theta _{\mathbf{R}}\right) e^{im\phi _{\mathbf{R}}}{\cal F}_{l}\left( \beta ,\left| \mathbf{R}\right| 
\right) 
\nonumber\\
&+&\frac{4\pi ^{\frac{3}{2}}(\frac{i}{2})^{l}}{V_{\rm UC}\Gamma \left( l+\frac{1}{2}\right) }
\nonumber\\
&&\sum _{\mathbf{G}\neq \left\{ \emptyset \right\} }\, \left| \mathbf{G}\right| ^{l-2}e^{-\frac{\pi ^{2}\left|
 \mathbf{G}\right| ^{2}}{\beta ^{2}}}\widetilde{P_{l}}^{m}\left( \cos \theta _{\mathbf{G}}
\right) e^{im\phi _{\mathbf{G}}} \, ,
\nonumber\\
\label{C4}
\end{eqnarray}
where $\mathbf{G}$ are reciprocal lattice vectors.
With an appropriate choice of \( \beta \sim \sqrt{\pi }/\left( V_{\rm UC}\right) ^{\frac{1}{3}} \),
and summing terms from smallest to largest, the periodic multipole interaction tensor can be computed
to high precision assuming an accurate representation of the incomplete gamma function.  In 
previous work by CWHG, the upward recursion
\begin{equation}
\label{C5}
\Gamma \left( m+1,\, x\right) =m\Gamma \left( m,\, x\right) +x^{m}e^{-x}
\end{equation}
was used,  which results in a loss of precision for large values of \( x \) and \( m \),
demanding extended precision arithmetic and precluding on the fly computation.
This problem is overcome by analytically summing the gamma function, collecting terms, and then
rewriting it  as
\begin{eqnarray}
\Gamma \left( m+\frac{1}{2},\, x\right) =\Gamma \left( m+\frac{1}{2}\right) 
\qquad\qquad\qquad
\nonumber\\
\left\{ {\rm erfc}\left( \sqrt{x}\right) 
+\sqrt{\frac{x}{\pi }}\sum _{n=0}^{m-1}(S_{n}\, x\, e^{-\frac{x}{n}})^{n} \right\} \, ,
\label{C7}
\end{eqnarray}
where the terms 
\begin{equation}
\label{SN}
S_{n}=\left( \frac{\Gamma \left( \frac{1}{2}\right) }{\Gamma \left( n+\frac{3}{2}\right) }
\right) ^{\frac{1}{{\rm max}(n,1)}} \, ,
\end{equation}
are simply pretabulated. This version of the gamma function is both easy to program and precise, even
for large values of \( x \) or \( m \). 

In one dimension, the \( {\cal M} \) tensor can be computed analytically as 
\begin{eqnarray}
{\cal M}^{m}_{l} & = & 
\frac{\widetilde{P}_{l}^{m}\left( \cos \theta _{0} \right) e^{im\phi _{0}}}{a_{0}^{l+1}}
\sum _{n=n_{0}}^{\infty }\frac{1}{n^{l+1}}
\nonumber\\
&+&\frac{\widetilde{P}_{l}^{m}\left( \cos \left( \theta _{0}+\frac{\pi }{2}\right) \right) 
e^{im\left( \phi _{0}+\pi \right) }}{a_{0}^{l+1}}
\sum _{n=n_{0}}^{\infty }\frac{1}{n^{l+1}}
\nonumber\\
& = & Q_{l}^{m}\left[ a_{0},\theta _{0},\phi _{0}\right] \left\{ \zeta (l+1)-
\sum _{n=1}^{n_{0}-1}\frac{1}{n^{l+1}}
\right\}
\nonumber\\
\label{C9}
\end{eqnarray}
where \( a_{0} \), \( \theta _{0} \) and \( \phi _{0} \) are the
initial box dimension and angles which are independent of the summation, and 
$\zeta$ is the Riemann zeta function \cite{MAbramowitz87}.

In two dimensions, the Fourier integrals for the calculation of the \( {\cal M} \) tensor
become more complicated. Taking the limit as the box dimension in the non-periodic
direction  goes to infinity (the $z$ direction in the following), we obtain from Eq.~\ref{C4} 
\begin{eqnarray}
{\cal M}^{m}_{l} 
 & = & \sum _{{\mathbf{R}}\in V_{\rm PFF}}\, \left( l-m\right) !\, P_{l}^{m}\left( \cos 
\theta _{\mathbf{R}} \right) e^{im\phi _{\mathbf{R}}}\, {\cal G}_{l}\left( \beta ,\left| 
\mathbf{R}\right| \right) 
\nonumber\\
 & - & \sum _{{\mathbf{R}'}\in V_{\rm In}}\, \left( l-m\right) !\, P_{l}^{m}\left( \cos  \theta _{
\mathbf{R}}\right) e^{im\phi _{\mathbf{R}}}{\cal F}_{l}\left( \beta ,\left| \mathbf{R}\right| \right)
\nonumber\\
 & + & {\cal C}_2 \sum _{\mathbf{G}\neq \left\{ \emptyset \right\} }\, \int _{-\infty }^{\infty }
\: dG_{z}\: \left| \mathbf{G}
\right| ^{l-2}e^{-\frac{\pi ^{2}\left| \mathbf{G}\right| ^{2}}{\beta ^{2}}}\, 
\nonumber\\
&& \qquad \qquad \qquad \qquad  {\tilde P}_{l}^{m}
\left( \cos  \theta _{\mathbf{G}} \right) e^{im\phi _{\mathbf{G}}} \, ,
\nonumber\\
\label{C10}
\end{eqnarray}
where
\begin{eqnarray}
{\cal C}_2 =  \frac{4\pi ^{\frac{3}{2}}(\frac{i}{2})^{l}}{A_{\rm UC}\Gamma \left( l+\frac{1}{2}\right) }
\label{C10p}
\end{eqnarray}
and \( A_{\rm UC} \) is the area of the cell along the non-periodic
direction. In practice, we carry out numerical evaluation of this integral.

\end{document}